# Intrinsic Even-Odd Thickness-Driven Anomalous Hall in Epitaxial MnBi$_2$Te$_4$ Thin Films

Debarghya Mallick[1], Simon Kim[1], An-Hsi Chen[1], Gabriel A. Vázquez-Lizardi[2], Alessandro R. Mazza[3], T. Zac Ward[4], Gyula Eres[1], Yue Cao[5], Debangshu Mukherjee[6], Hu Miao[1], Liang Wu[7], Christopher Nelson[4], Danielle Reifsnyder Hickey[2,8,9], Robert G. Moore[1], Matthew Brahlek[1*]

[1]Materials Science and Technology Division, Oak Ridge National Laboratory, Oak Ridge, TN, 37831, USA
[2]Department of Chemistry, Pennsylvania State University, University Park, PA 16802, USA
[3]Materials Science and Technology Division, Los Alamos National Laboratory, Los Alamos, NM 87545, USA
[4]Center for Nanophase Materials Science, Oak Ridge National Laboratory, Oak Ridge, TN, 37831, USA
[5]Materials Science Division, Argonne National Laboratory, Lemont, IL 60439, USA
[6]Computing and Computational Sciences Directorate, Oak Ridge National Laboratory, Oak Ridge, TN, 37831, USA
[7]Department of Physics and Astronomy, University of Pennsylvania, Philadelphia, PA, 19104
[8]Department of Materials Science and Engineering, Pennsylvania State University, University Park, PA 16802, USA
[9]Materials Research Institute, Pennsylvania State University, University Park, PA 16802, USA

Correspondence should be addressed to *brahlekm@ornl.gov

**Abstract**: We demonstrate precise control of magnetism in MnBi$_2$Te$_4$ thin films through careful synthesis by molecular beam epitaxy, achieving minimal defects and accurate layer thickness control. By optimizing Mn-Bi-Te ratios and growth temperatures, we minimize detrimental self-doping effects and accurately target integer-layer films. X-ray diffraction and reflectivity provide quantitative measures of film quality and thickness. When these macroscale probes of structure and thickness are integrated with magnetotransport measurements, a striking even-odd layer dependence of the anomalous Hall effect is revealed. Odd-layer films exhibit a large hysteresis up to the Néel temperature (~25K), consistent with non-compensated antiferromagnetism, while even-layer films show minimal response, as expected for an antiferromagnet. The sign of the anomalous Hall effect exhibits a sign reversal for intrinsic magnetism versus magnetism associated with defects. This work identifies critical factors for inducing pure, non-compensated ferromagnetism and reveals the character of the intrinsic anomalous Hall effect in MnBi$_2$Te$_4$, which together is a step towards realizing the zero-field quantum anomalous Hall effect in topological materials.

PhySH terms: Topological insulators, Chern insulators, Anomalous Hall effect, Quantum anomalous Hall effect, Antiferromagnets, X-ray diffraction, Molecular beam epitaxy, Resistivity measurements

This manuscript has been authored by UT-Battelle, LLC under Contract No. DE-AC05-00OR22725 with the U.S. Department of Energy. The United States Government retains and the publisher, by accepting the article for publication, acknowledges that the United States Government retains a non-exclusive, paid-up, irrevocable, world-wide license to publish or reproduce the published form of this manuscript, or allow others to do so, for United States Government purposes. The Department of Energy will provide public access to these results of federally sponsored research in accordance with the DOE Public Access Plan (http://energy.gov/downloads/doe-public-access-plan). [DOES NOT NEED TO BE INCLUDE IN PUBLISHED VERSION]

The quantum anomalous Hall effect emerges when time reversal symmetry is broken in a topological insulator, giving rise to emergent 1D chiral edge states at the boundary. Thus, the Hall resistance is exactly quantized at zero magnetic field [1–4]. Gaining control over the quantum anomalous Hall effect requires gapping topological Dirac states via internal magnetism [3,4]. Several routes have been explored including interfacing topological insulators with magnetic materials at atomically precise interfaces [5], magnetically doping topological insulators [6–8], and looking for materials that are simultaneously both magnetic and topological. The field of magnetic topological materials has primarily focused on Cr and V doping the tetradymite topological insulator family $Bi_2Se_3$, $Bi_2Te_3$, $Sb_2Te_3$ and mixtures. Cr and V doped $(Bi,Sb)_2Te_3$ has been the work-horse material since the Sb to Bi ratio can be adjusted for charge defect compensation which enables routes to control the location of the Fermi energy such that it is in the Dirac gap, thus exhibiting quantized transport [8]. Despite the large amount of success seen in this area, there remain clear challenges, which includes low ordering temperature (~10 K) and a small net moment that results in a small gap, and thus low temperature for perfect quantization. Moreover, the Dirac point lies close to the valence band, and the magnetic and compensation doping both result in large levels of disorder. These challenges highlight the difficulties in identifying materials that are both intrinsically magnetic as well as topological.

The most prominent candidate that is both intrinsically topological and magnetic has been the Mn-Bi-Te homologous family of materials [9–11]. This series of materials derives from the tetradymite quintuple layer (QL) unit, Te-Bi-Te-Bi-Te, which is split at the center Te layer and a 2D layer of Mn-Te is inserted to form a septuple layer (SL) unit, Te-Bi-Te-Mn-Te-Bi-Te, as shown in Fig. 1(a). Higher-order members are then obtained by layering a single SL with n-layers of QL, for example $n = 1$ is composed of a QL of $Bi_2Te_3$ and a SL of $MnBi_2Te_4$ yielding $MnBi_4Te_7$ [12]. The $n = 0$ member (the focus of the current work), $MnBi_2Te_4$ is both topological with a single Dirac cone on the surface as well as being antiferromagnetic with $T_N \approx 25K$ and a large net moment per unit cell [13–15]. As shown in Fig. 1(b), the antiferromagnetic structure is composed of ferromagnetic layers that are aligned in opposite directions. For thin films, this magnetic structure then opens a novel route to induce a net moment by employing the uncompensated moment that arises for an odd number of layers, which contrasts even number of layers that should possess no net moment, as shown in Fig. 1(b). This finite-thickness route to creating ferromagnetism is thus an exciting mechanism to stabilize the quantum anomalous Hall effect at zero field and high temperature [16]. This early success was, however, tempered by the fact that the Mn-Bi-Te system is both highly prone and extremely sensitive to defects [17,18]. For example, as highlighted in the high-angle annular dark-field scanning transmission electron microscopy (HAADF-STEM) image in Fig. 1(c-d) $MnBi_2Te_4$ exhibits antisite defects (Bi on Mn site, Fig. 1(d)), vacancies, and QL intergrowth (Fig. 1(c)) that can result from either excess Bi or from non-ideal growth temperatures where the Mn is incorporated into a QL intergrowth rather than forming the targeted SL. All defects have profound implications on the electronic and magnetic properties of these materials. Thus, understanding how to grow samples where the defects are minimized and the thickness is accurately controlled remains a challenge.

In this work, we utilize molecular beam epitaxy (MBE) to grow a series of $MnBi_2Te_4$ with minimal defects and accurately controlled thickness to realize fully non-compensated magnetism. This is achieved by carefully adjusting the Mn-Bi-Te ratios and temperatures during growth, which minimizes deleterious self-doping effects that give rise to ferromagnetism well below $T_N$. The magneto-transport response of these films then reveals an anomalous Hall effect that is nominally absent in even-layer films and is very large in odd-layer films. Moreover, in contrast to films with intentional impurities where the ferromagnetic signature vanishes near $T \approx 10$-15 K, the anomalous Hall effect follows the antiferromagnetic transition,

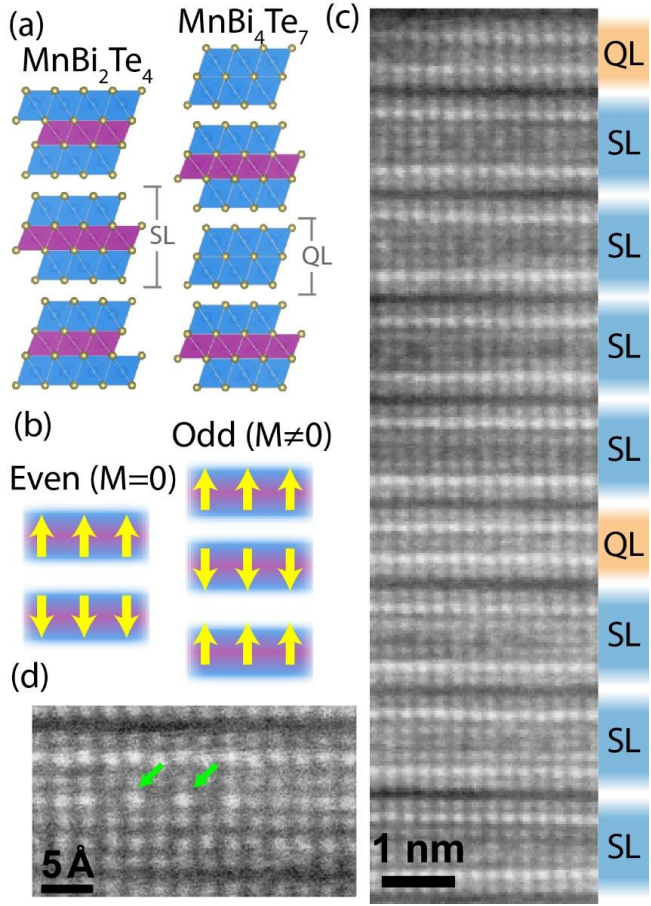

**Fig. 1** (a) Schematic of the MnBi$_2$Te$_4$ and the MnBi$_4$Te$_7$ structures, where MnBi$_2$Te$_4$ is composed of septuple layers (SLs, blue and purple layers) and MnBi$_4$Te$_7$ is composed of alternating SLs and quintuple layers (QLs, blue layers). (b) Schematic of the layered antiferromagnetic ordering in even and odd layers of MnBi$_2$Te$_4$. (c-d) HAADF-STEM images of a MnBi$_2$Te$_4$ film grown with a slight excess of Bi, where QL intergrowths were present, as highlighted in (c), as well as antisite defects where excess Bi appears on the Mn sites, as indicated by the green arrows (d).

which confirms the non-compensated origins. Combining transport, which is sensitive to QL intergrowths and antisite defects, with X-ray measurements proves an effective and simple method to optimize the film growth. Here, diffraction is quantitatively very sensitive to QL intergrowths in the thin-limit, and film thickness is accurately measured by X-ray reflectivity, which together with transport provides unambiguous quantitative feedback on the growth, thus enabling targeting integer layer films with intrinsic properties. This work is the first step toward understanding critical factors required to leverage finite thickness control of films, which is necessary for inducing the magnetism needed to realize zero-field quantum anomalous Hall effect.

The samples used in this study were grown using MBE as described in Ref. [19,20] for thick samples. The main advancement in this work is the integration of quantitative X-ray analysis into the growth process, enabling precise control of both film structure and thickness to achieve an exact integer number of layers with high structural quality. The flux for the Bi, Mn, and Te cells were carefully calibrated using an in situ quartz crystal microbalance that was cross calibrated with X-ray reflectivity and Rutherford backscattering. The flux ratio used for the samples was kept between the Bi:Mn ratio of 2.0-2.6, and the Te was maintained at greater than 5-10 times the Bi flux. The MnBi$_2$Te$_4$ samples were grown on Al$_2$O$_3$ (0001) substrates, which were mounted to the sample platens with silver paste and cured ex situ. Once in the MBE, the substrates were heated to 600 °C prior to growth. The films were grown using a two-step method where a 3-SL nucleation layer was grown at low temperature and subsequently heated to 225 °C where the remaining film was deposited to attain the targeted thickness. The samples were grown by codepositing Mn, Bi, and Te. After growth, all samples used for magnetotransport were capped in situ with Te.

X-ray diffraction and reflectivity are very sensitive indicators of both the phase purity of MnBi$_2$Te$_4$ and the thickness, and they were used to rapidly identify non-idealities in the growth and provide very accurate thickness control. The HAADF-STEM image in Fig. 1(c) shows a sample grown intentionally with excess Bi, which highlights one of the most common defects, a QL intergrowth. This defect indicates either excess Bi supplied to the growing surface or from the growth temperature being too low and nucleation of the QL is favored. The temperature can be optimized by monitoring the expected peak positions for MnTe impurities, and then the optimized value is just below the temperature where MnTe is no longer resolved, as described in [19]. Following this, the Mn to Bi ratio can be adjusted such that ideal stoichiometry and phase purity are achieved. It was found that there are sensitive indicators in X-ray diffraction of the formation of QL intergrowths, especially for thin samples. To show this quantitatively, we simulated the specular X-ray diffraction with a kinematic model composed of an $n$-SL $MnBi_2Te_4$ film with a single QL intergrowth on a $Al_2O_3$ substrate, as detailed in [21]. As shown in the schematic in Fig. 2(a), the models were calculated with the QL intergrowth at all possible positions and the subsequent intensities averaged, and the results are shown in Fig. 2(a) for thickness between 3SL and 48 SL. The diffraction curves show splitting for certain peaks, which indicate the intergrowths are present. For example, the 006 peak shows a clear splitting of about 3° for the 3 SL sample and decreases with increasing thickness. This reduction in the splitting at larger thicknesses occurs because the effective concentration of the intergrowths decreases and the peak width reduces with increasing thickness. For larger concentrations of QLs the evolution of the peak splitting can be further seen in the simulations in the supplement [21] and in previous works Ref [19], where the splitting systematically increases and smoothly merges into the longer-range order for the $MnBi_4Te_7$.

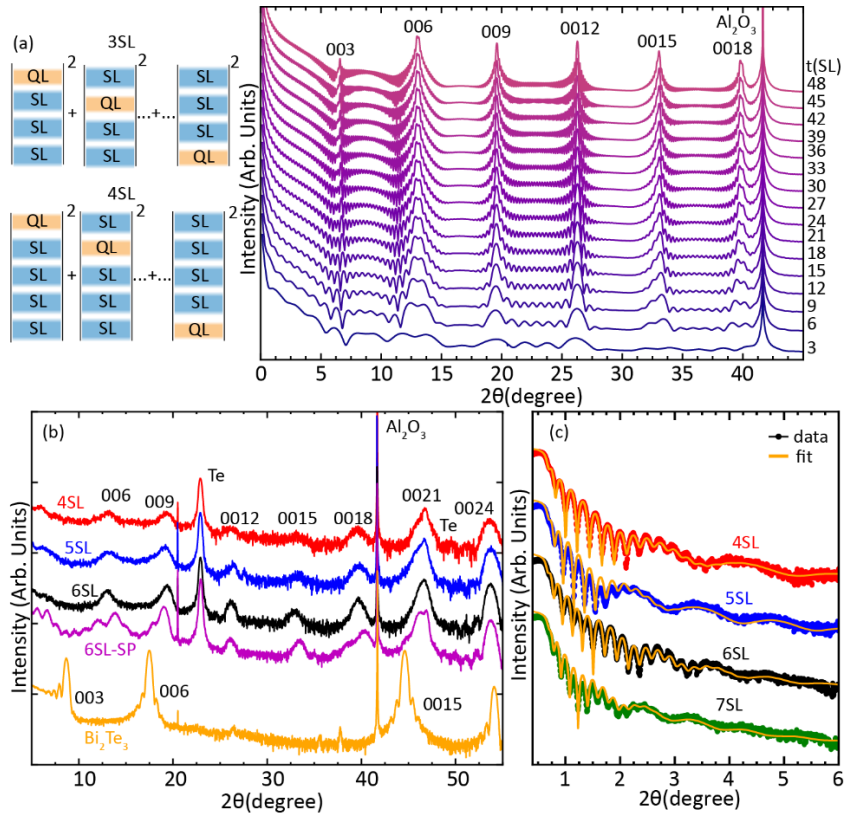

These simulations give a guide for optimizing how $MnBi_2Te_4$ was grown and benchmarked against experimental X-ray measurements performed on a 4-circle Panalytical MRD using Cu $k_{\alpha 1}$ radiation. X-ray diffraction $2\theta$-$\theta$ curves are shown in Fig. 2(b) for several samples that demonstrate this process where the relative Mn to Bi was systematically reduced until the 006 peak was a clear single peak. From the top, there are 4-6 SL samples that were grown with this method where the $MnBi_2Te_4$ 00$L$ peaks are labeled, as well as the Te capping layer and the peaks from the $Al_2O_3$ substrate. The 4 SL, 5 SL

**Fig. 2.** (a) Simulated X-ray diffraction curves for $MnBi_2Te_4$ samples with QL intergrowths, as shown in the schematic at the left. (b) X-ray diffraction $2\theta$-$\theta$ scans of stoichiometric $MnBi_2Te_4$ and one sample grown with excess Bi, as well as a reference $Bi_2Te_3$ sample. (c) X-ray reflectivity of stoichiometric $MnBi_2Te_4$ samples. All curves are logarithms of the intensity.

**Table 1.** Fitting parameters for XRR data shown in Fig. 2(c).

| | MnBi$_2$Te$_4$ film | | | | | | Te cap | | |
|---|---|---|---|---|---|---|---|---|---|
| Target (SL) | Thickness (Å) | RMS (Å) | Thickness (SL) | Δ (SL) | RMS (SL) | Density (g/cm$^3$) | Thickness (Å) | RMS (Å) | Density (g/cm$^3$) |
| 4 | 52.0 | 1.0 | 4.0 | -0.2 | 0.1 | 7.361 | 297 | 10 | 6.243 |
| 5 | 64.0 | 1.0 | 5.0 | -0.2 | 0.1 | 7.361 | 317 | 14 | 6.243 |
| 6 | 84.0 | 3.0 | 6.0 | 0.2 | 0.2 | 7.361 | 305 | 10 | 6.243 |
| 7 | 94.0 | 4.0 | 7.0 | -0.1 | 0.3 | 7.361 | 335 | 16 | 6.243 |

and 6 SL samples all show well-defined single peaks, indicating high-quality MnBi$_2$Te$_4$. For comparison, a sample of thickness ~6 SL that was grown with intentional excess Bi is shown, where the 006 peak exhibits a large splitting, consistent with the simulations. Moreover, the 009 peak of this film exhibits an additional small shoulder at lower 2$\theta$. This split peak is absent in the simulation, which suggests that it is likely additional QL phase segregation and a proto-peak for Bi$_2$Te$_3$. This can be seen by comparing the Bi$_2$Te$_3$ curve below, where the 006 peak aligns with the shoulder and indicates that areas of this film may contain a small amount of intergrowth of 2 or more QLs, which can occur if the stoichiometry is far off or if the growth temperature is too low. Sample-to-sample variations under nominally identical growth conditions were observed as formation of MnBi$_2$Te$_4$ or Mn-doped Bi$_2$Te$_3$ and were most likely the result of slight temperature variations during formation. This emphasizes the very narrow temperature window that separates Mn-doped Bi$_2$Te$_3$ when the temperature is too low, MnBi$_2$Te$_4$ at ideal temperatures, and MnBi$_2$Te$_4$ with precipitated MnTe when the temperature is too high [19,22,23]

Once the stoichiometry and phase are confirmed, X-ray reflectivity was used to quantify the overall growth rate, thereby making it possible to target an integer number of layers. X-ray reflectivity is a powerful technique for probing thin-film thickness (with sub-angstrom accuracy), morphology, and density. It measures these properties on a macroscopic scale (> mm$^2$), matching the length scale relevant for transport measurements. This sensitivity arises from coherent scattering at film interfaces, producing oscillations that decay strongly in 2$\theta$–$\theta$ scans. The real-space thickness profile can be approximated as the Fourier transform of the reflectivity curve but can be accurately modeled using established methods [24]. Shown in Fig. 2(c) are specular reflectivity curves for 4 SL, 5 SL, 6 SL and 7 SL samples that were capped by Te. The data exhibits two clear sets of oscillations. The high-frequency oscillations that are more prominent at low 2$\theta$ derive from the thicker Te capping layer, and the lower-frequency oscillations that are more apparent at high 2$\theta$ are due to the MnBi$_2$Te$_4$ film. The data were fit using the GenX program [23] using a two-layer model (MnBi$_2$Te$_4$ and the Te cap), and the resulting curves are overlain on the data. As compiled in Tab. 1, the resulting parameters gave a thickness of about 30 nm and a roughness of 1 nm of the Te cap, and the MnBi$_2$Te$_4$ thickness of 4 SL, 5 SL, 6 SL and 7 SL with an error within 0.1-0.2 SL, similar to the roughness of 0.1 SL, 0.1 SL, 0.2 SL and 0.3 SL, respectively. The densities of Al$_2$O$_3$ and MnBi$_2$Te$_4$ were taken to be the bulk values and Te was taken to be a fitting parameter but was within a few percent of the bulk value, possibly due to the formation of a surface oxide on the Te cap. Goodness of fit metrics such as $\chi^2$ were not useful since experimental artifacts at low 2$\theta$ (beam footprint effects, slight misalignments and air scattering effects) are not captured in the model. Since, these deviations occur in high intensity regions, $\chi^2$ is artificially large. However, the accuracy in thickness extracted from X-ray reflectivity is evident in Fig. S2, where the thickness varied by sub-SL increments and the resulting curved showed a strong disagreement with the data. In situ metrologies such as RHEED oscillations were not viable to quantify the growth rate since oscillations were not observed, which indicates that the growth mode is likely step-flow, like $Bi_2Se_3$ [25]. As such, prior to the growth of this set of samples, calibration samples were grown that resulted in non-integer thicknesses in SL, and the growth rate was then adjusted until the thickness was an integer value up

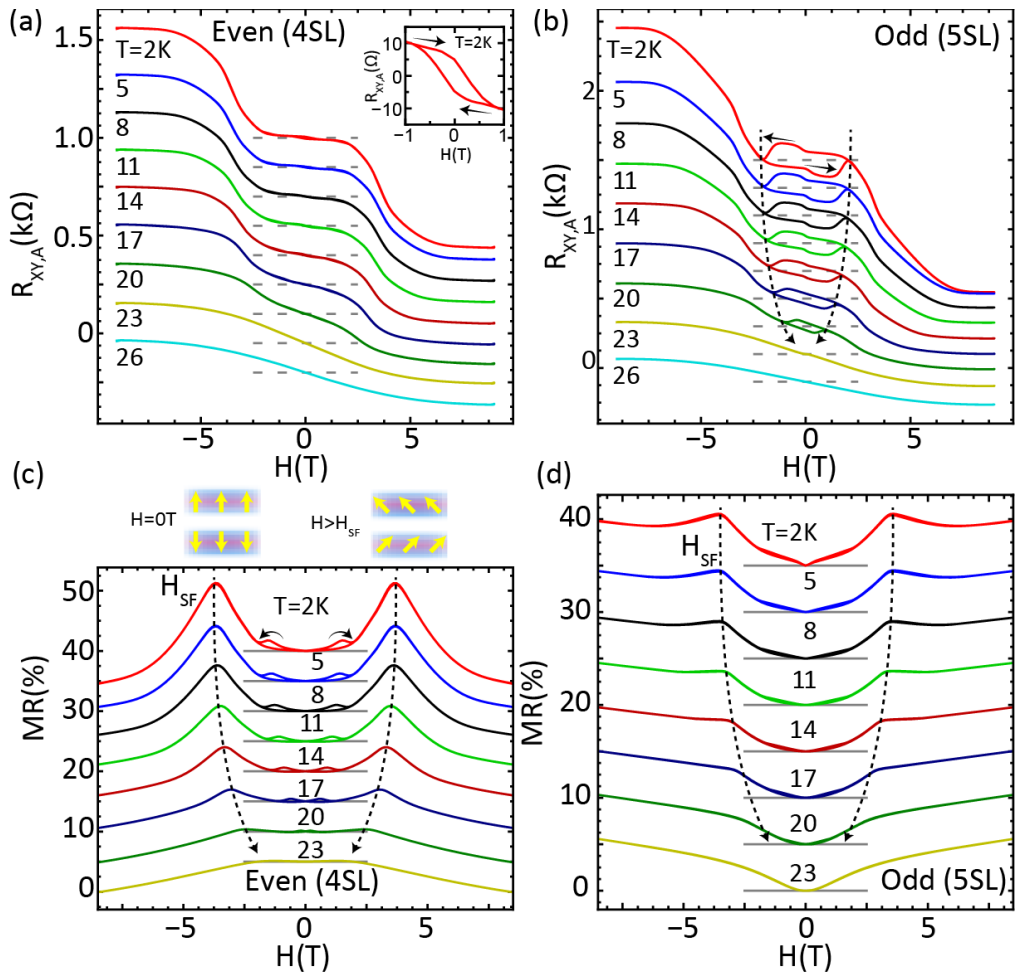

**Fig. 3.** (a-d) Temperature dependence of the anomalous Hall effect (a,b) and magnetoresistance (c,d) versus temperature for even (4SL, a,c) and odd (5SL, b,d). The schematic at the top of (c) highlights the flopping of the anisotropy direction with an out-of-plane applied magnetic field for $MnBi_2Te_4$. The small arrows in (b,c) and in the inset in (a, zoom in of the 2K data) indicate the sweep direction.

to an accuracy of the typical roughness.

Temperature-dependent magnetotransport measurements are shown for an even-thickness sample (4 SL) and an odd-layer sample (5 SL) in Fig. 3. The measurements were performed using pressed indium wires contacted in van der Pauw geometry, in a field up to 9 T applied orthogonal to the films' surface, and down to a base temperature of 2 K. The data was antisymmetrized to remove unavoidable mixing of $R_{xx}$ and $R_{xy}$, as shown in [21]. The anomalous Hall effect ($R_{xy,A}$, the Hall effect with a linear background removed, as shown in [21]), is shown in Fig. 3(a, even) and (b, odd), and the magnetoresistance $MR=(R(H)-$ $R(H=0\text{T}))/R(H=0\text{T}))\times 100\%$. These two quantities give insight into the field-induced dynamics of any emergent ferromagnetism as well as the ground-state antiferromagnetism. All the data in Fig. 3 were offset for clarity, the horizontal line indicates zero. First, contrasting with the anomalous Hall effect for the samples with even and odd thicknesses, the even-layer sample exhibits virtually no hysteresis at 2 K compared to the very large hysteresis of the odd-layer sample [26–32]. Moreover, the curves taken at higher temperature indicate the nominal transition temperature: as highlighted by the dashed arrows, the hysteresis in the odd-layer sample clearly vanishes between 23-26 K, which is consistent with the bulk antiferromagnetic behavior near 25 K [15,30,33,34].

The magnetoresistance curves shown in Fig. 3(c-d) exhibit several characteristics that indicate the antiferromagnetic properties, which are nominally similar for both the even- and the odd-SL samples [14,31]. For the low-field regime, the MR curves at 2 K show a monotonic increase with increasing magnetic field, which transitions to prominent peaks near 3-4 T. This peak has been identified with the spin-flop transition ($H_{SF}$), where the spin-alignment rearranges from out-of-plane to in-plane, as shown schematically at the top of Fig. 3(c) [27,35–37]. For fields larger in magnitude than $H_{SF}$, the MR decreases, as the canting of the spins increases with increasing field, which reduces the effective disorder and correspondingly the resistance decreases [31,38]. The temperature dependence of the spin-flop transition is indicated by the dashed arrows for both the even- and the odd-layer samples. For both samples, the transition clearly persists to around 20-23 K. Around this temperature range, the MR has a simple monotonic reduction with field, which indicates that the sample is in a paramagnetic state where field-polarization decreases the net scattering and thus the resistance decreases. Interestingly, the data show two slight differences among the samples. First, the 5 SL sample shows a net flatter MR compared to the 4 SL sample with a net positive MR at high field, whereas the MR is net negative for the 4 SL sample. This difference likely indicates that there are different levels of disorder in the 5 SL and 4 SL samples, which affect both the semiclassical orbital effect (positive MR) and negative MR associated with the evolution of magnetic domains with changing applied field. Both these effects are very sensitive to disorder effects, and, as such, deconvoluting the fundamental origins is challenging. Finally, the 4 SL sample exhibits slight hysteresis near the spin-flop transition, which closes around ±1 T. Since there is no hysteresis about zero field, this likely originates from nucleation and growth of the flopped antiferromagnetic domains, as has been previously reported [19].

Figure 4 summarizes the magnetotransport results for the series of samples grown with carefully controlled stoichiometry and thickness. Figure 4(a) shows a comparison of the anomalous Hall effect at 2 K. Here the data clearly shows that the odd-layer samples (5 and 7 SL) have very large hysteresis, whereas the even-layer samples (4 and 6 SL) exhibit almost no hysteresis. Figure 4(b) shows $\Delta R_{XY,A} = R_{XY,A}(H_\uparrow) - R_{XY,A}(H_\downarrow)$, which is the difference between the up sweeps ($H_\uparrow$) and the down sweeps ($H_\downarrow$) for the data in Fig. 4(a). These data together highlight the large difference between the odd-layer samples and the even-layer samples. The character of the hysteresis gives additional insight regarding the fundamental physics of the anomalous Hall that originates from ferromagnetism related to the uncompensated moment, as well as the defects that exist in these samples. The hysteresis for the 5 and 7 SL samples show two features. The large hysteresis that closes near 2 T is associated with the antiferromagnetic phase, and 2 T correlates well with the field where the spin-flop onsets (Fig. 3(c-d)) [36]. A small depression can be seen in the odd-layer samples near zero field. This low-field hysteresis, often called *wasp-waisted hysteresis*, implies that there are multiple magnetic components with different coercivities, one associated with the non-compensated moment and the second soft component likely associated with defects (antisite defects or small concentrations of QL intergrowths).

The full field- and temperature-dependent data in Fig. 3(b) shows that the two components of hysteresis for the odd-layer samples have two temperature dependences [22,38,41]. The dominant component vanishes near 23 K, whereas the soft component associated with the depression vanishes near 14-17 K. This can be more clearly seen by plotting the temperature dependence of the $\Delta R_{XY,A}$ data, which are summarized in Fig. 4(c). The data in Fig. 4(c) shows $\Delta R_{XY,A}$ taken at zero field (solid curves with symbols) and at 0.5 T (gray curves, no symbols). Plotted vertically below is the spin-flop field versus temperature, which vanishes near the bulk Neel temperature around 25 K. First, $\Delta R_{XY,A}(H=0T)$ for the even-layer samples is more than an order of magnitude smaller than for the odd-layer samples, which is more clearly shown in the zoom-in for the even-layer samples in the inset of Fig. 4(c). For both the odd- and the even-layer samples, $\Delta R_{XY,A}(H=0T)$ vanishes near 23 K, which is fully consistent with the Neel temperature, as measured by the vanishing of the spin-flop field. The vanishing of this transition with increasing temperature is shown in Fig. 4(d) for all samples. These data show that the transition varies little with thickness and all samples $H_{SF}$ disappears near 23 K, consistent with the temperature where hysteresis in the anomalous Hall effect goes away. This confirms that the anomalous Hall effect originates from the

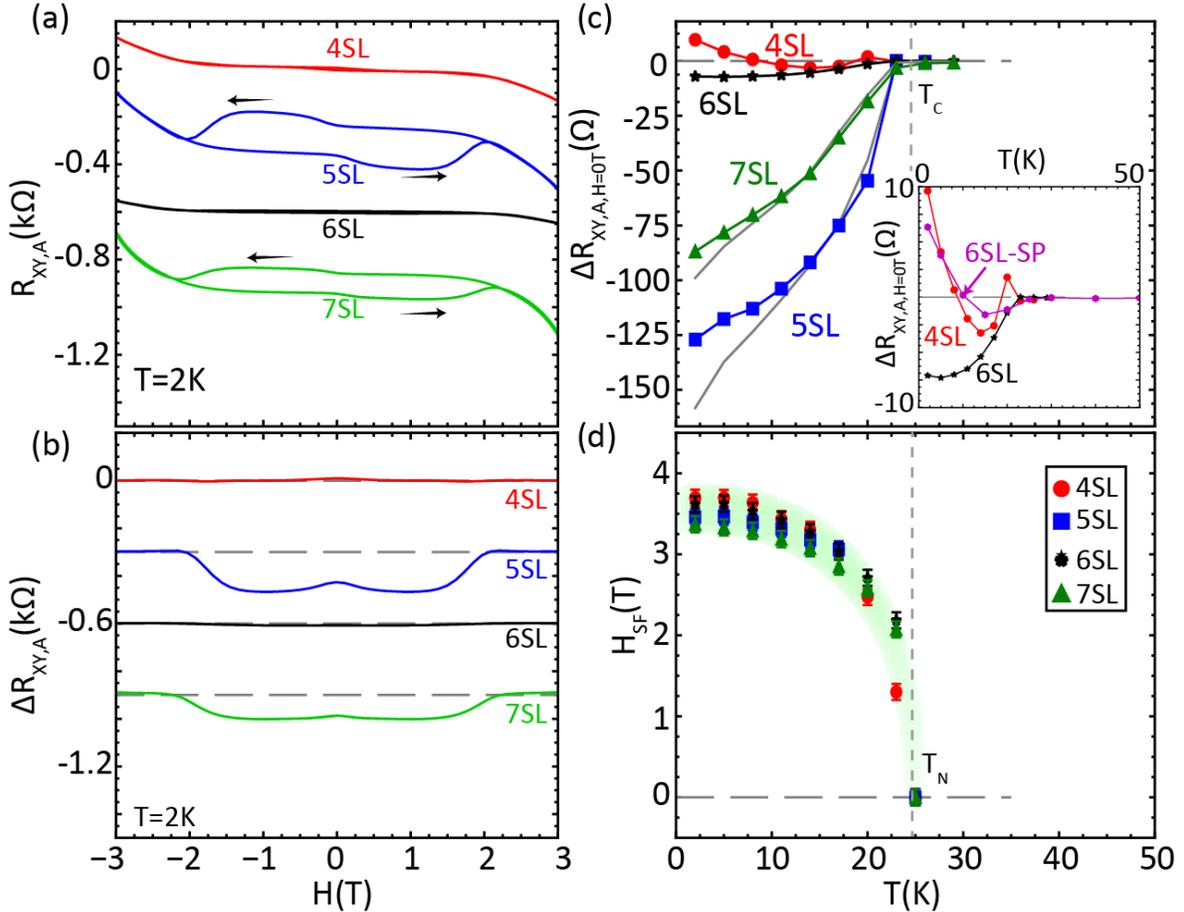

**Fig. 4** (a) Anomalous Hall effect at a temperature of 2K for 4, 5, 6, 7 SL. Small arrows indicate sweep direction. (b) Difference of anomalous Hall effect in (a) taken during the up-sweep and down-sweep of the magnetic fields. (c) Temperature dependence of the zero-field anomalous Hall effect for 4, 5, 6, 7SL. The solid gray curves were taken at 0.5 T. The inset shows a zoom-in of the 4 and 6 SL samples, as well as a 6 SL sample grown with excess Bi with a split peak (SP) shown in Fig. 2. (d) Temperature dependence of the spin-flop (SF) field taken from the data in Fig. 3.

antiferromagnetism and is consistent with the more than an order-of-magnitude larger $\Delta R_{XY,A}(H=0T)$ for odd- compared to the even-thickness films.

The prominent even-odd thickness change of the MnBi$_2$Te$_4$ films highlights the effectiveness of careful defect and thickness control. These data enable clear identification of the anomalous Hall effect in both the even- and odd-layer samples, which can be resolved through the temperature dependence and the sign of $\Delta R_{XY,A}(H=0T)$. First, the odd-layer samples show a large and negative $\Delta R_{XY,A}(H=0T)$ sign, which contrasts with the analysis of previous studies which concluded that the sign is opposite for even vs odd layer samples [38]. The sign of the AHE is found to be the same for the 6 SL sample. This confirms that the intrinsic sign of the anomalous Hall effect related to the antiferromagnetism is negative in our samples [38]. The 4 SL sample at 2 K shows a tiny $\Delta R_{XY,A}(H=0T)$ with a positive sign, which changes sign with increasing temperature near 9 K, before decreasing to zero near 23 K. This sign change implies two origins for the anomalous Hall effect, consistent with the odd-layer samples that have a soft component associated with defects and a harder component associated with antiferromagnetism. To quantify the difference, Fig. 4(c) shows $\Delta R_{XY,A}(H=0T)$ and $\Delta R_{XY,A}(H=0.5T)$ for the odd-layer samples. $\Delta R_{XY,A}(H=0.5T)$ captures only the anomalous Hall from antiferromagnetism, whereas $\Delta R_{XY,A}(H=0T)$ is sensitive to the ferromagnetism associated with the non-compensated moment and the soft impurity. These two sets of curves clearly merge near 10-12 K, which is consistent with the sign change of the 4 SL sample and previous studies that show that defect-induced ferromagnetism has a $T_C$ near 10-15 K, well below the Neel temperature [38]. Furthermore, the anomalous Hall curve for a 6 SL sample that was grown with additional Bi (the sample whose X-ray diffraction is shown in Fig. 2) shows that $\Delta R_{XY,A}(H=0T)$ is similar to that shown for the 4 SL, confirming that the origin is related to QL intergrowths. Finally, the temperature dependence of $\Delta R_{XY,A}(H=0T)$ for the 6 SL is qualitatively identical to the 5 SL and 7 SL samples, with the only difference being the maximum magnitude of $\Delta R_{XY,A}$. This implies that the origin of the anomalous Hall effect is the same as for the odd-thickness samples. The cause must be derived from the thickness and can be the result of either (1) the 6 SL sample is slightly off from the targeted thickness of 6 SL, for example, 90% of the sample is 6 SL and the remaining 10% is either 5 or 7 SL; or (2) there may be islands of 7 SL and depressions of 5 SL, resulting in a net 6 SL sample, due to the island formation common to these materials during growth [29,41–43]. Both scenarios are consistent with the RMS roughness from the reflectivity model. An interesting future direction would be correlating direct measurements of surface morphology and roughness by in situ scanning probe microscopy prior to sample capping and ex situ measurements by reflectivity and transport. These experiments would give a better understanding of the fundamental factors (islands/hole formation and interface properties) that determine the characteristics of the anomalous Hall effect.

To conclude, we have shown that with careful defect and thickness control during synthesis it is possible to produce MnBi$_2$Te$_4$ films with intrinsic even and odd SL-dependent properties. We have shown that this can be achieved through cross-referencing growth conditions with X-ray diffraction and reflectivity which are extremely sensitive metrics of the macroscale structural and thickness. The resulting high-quality films with accurate even and odd SL thicknesses enable understanding functionally coupled electronic and magnetic responses. Here, the odd-SL samples show a large anomalous Hall effect whereas this effect is nearly absent in the even-SL samples. This, along with the temperature dependence clearly ties the anomalous Hall effect to the antiferromagnetism for which the net moment is exactly compensated in even-layer samples and uncompensated for samples having an odd number of layers. Interestingly, the sign for $\Delta R_{XY,A}(H=0T)$ (up-sweep minus down-sweep) is negative for the odd-layer samples, indicating the intrinsic response, whereas previous studies suggest the sign is odd for a typical impurity found in MBE-grown

MnBi$_2$Te$_4$. This approach enables utilization of intrinsic magnetic topological insulator phases to more readily access the quantized anomalous Hall effect, and it may offer a route to large exchange gaps that will enable higher quantization temperature.

**References**
[1] M. Z. Hasan and C. L. Kane, Colloquium: Topological insulators, Rev. Mod. Phys. **82**, 3045 (2010).
[2] M. Z. Hasan and J. E. Moore, Three-Dimensional Topological Insulators, Annual Review of Condensed Matter Physics **2**, 55 (2011).
[3] C.-X. Liu, S.-C. Zhang, and X.-L. Qi, The Quantum Anomalous Hall Effect: Theory and Experiment, Annual Review of Condensed Matter Physics **7**, 301 (2016).
[4] C.-Z. Chang, C.-X. Liu, and A. H. MacDonald, *Colloquium* : Quantum anomalous Hall effect, Rev. Mod. Phys. **95**, 011002 (2023).
[5] R. Watanabe, R. Yoshimi, M. Kawamura, M. Mogi, A. Tsukazaki, X. Z. Yu, K. Nakajima, K. S. Takahashi, M. Kawasaki, and Y. Tokura, Quantum anomalous Hall effect driven by magnetic proximity coupling in all-telluride based heterostructure, Applied Physics Letters **115**, 102403 (2019).
[6] Y. Ou et al., Enhancing the Quantum Anomalous Hall Effect by Magnetic Codoping in a Topological Insulator, Advanced Materials **30**, 1703062 (2018).
[7] H. T. Yi, D. Jain, X. Yao, and S. Oh, Enhanced Quantum Anomalous Hall Effect with an Active Capping Layer, Nano Lett. **23**, 5673 (2023).
[8] C.-Z. Chang et al., Experimental Observation of the Quantum Anomalous Hall Effect in a Magnetic Topological Insulator, Science **340**, 167 (2013).
[9] M. M. Otrokov et al., Highly-ordered wide bandgap materials for quantized anomalous Hall and magnetoelectric effects, 2D Mater. **4**, 025082 (2017).
[10] Y. Gong et al., Experimental Realization of an Intrinsic Magnetic Topological Insulator, Chinese Phys. Lett. **36**, 076801 (2019).
[11] J. Li, Y. Li, S. Du, Z. Wang, B.-L. Gu, S.-C. Zhang, K. He, W. Duan, and Y. Xu, Intrinsic magnetic topological insulators in van der Waals layered MnBi2Te4-family materials, Science Advances **5**, eaaw5685 (2019).
[12] K. He, MnBi2Te4-family intrinsic magnetic topological materials, Npj Quantum Mater. **5**, 1 (2020).
[13] Y.-J. Hao et al., Gapless Surface Dirac Cone in Antiferromagnetic Topological Insulator MnBi2Te4, Phys. Rev. X **9**, 041038 (2019).
[14] T. Hirahara et al., Fabrication of a novel magnetic topological heterostructure and temperature evolution of its massive Dirac cone, Nat Commun **11**, 4821 (2020).
[15] J.-Q. Yan, Q. Zhang, T. Heitmann, Z. Huang, K. Y. Chen, J.-G. Cheng, W. Wu, D. Vaknin, B. C. Sales, and R. J. McQueeney, Crystal growth and magnetic structure of MnBi2Te4, Phys. Rev. Mater. **3**, 064202 (2019).
[16] Y. Deng, Y. Yu, M. Z. Shi, Z. Guo, Z. Xu, J. Wang, X. H. Chen, and Y. Zhang, Quantum anomalous Hall effect in intrinsic magnetic topological insulator MnBi2Te4, Science **367**, 895 (2020).
[17] M. Garnica et al., Native point defects and their implications for the Dirac point gap at MnBi2Te4(0001), Npj Quantum Mater. **7**, 1 (2022).
[18] Z. Huang, M.-H. Du, J. Yan, and W. Wu, Native defects in antiferromagnetic topological insulator MnBi2Te4, Phys. Rev. Mater. **4**, 121202 (2020).
[19] J. Lapano et al., Adsorption-controlled growth of MnTe(Bi2Te3)n by molecular beam epitaxy exhibiting stoichiometry-controlled magnetism, Physical Review Materials **4**, 111201 (2020).

[20] A. R. Mazza et al., Surface-Driven Evolution of the Anomalous Hall Effect in Magnetic Topological Insulator MnBi2Te4 Thin Films, Advanced Functional Materials **32**, 2202234 (2022).
[21] supplementary material, (n.d.).
[22] K. Zhu, Y. Cheng, M. Liao, S. K. Chong, D. Zhang, K. He, K. L. Wang, K. Chang, and P. Deng, Unveiling the Anomalous Hall Response of the Magnetic Structure Changes in the Epitaxial MnBi2Te4 Films, Nano Lett. **24**, 2181 (2024).
[23] M. Björck, G. Andersson, M. K., B. D., S. J., P. W., W. M., and N. D., *GenX*: an extensible X-ray reflectivity refinement program utilizing differential evolution, Journal of Applied Crystallography **40**, 1174 (2007).
[24] S. Basu and S. Singh, *Neutron and X-Ray Reflectometry: Emerging Phenomena at Heterostructure Interfaces* (IOP Publishing, 2022).
[25] N. Bansal et al., Epitaxial growth of topological insulator Bi2Se3 film on Si(111) with atomically sharp interface, Thin Solid Films **520**, 224 (2011).
[26] C. Liu, Y. Wang, H. Li, Y. Wu, Y. Li, J. Li, K. He, Y. Xu, J. Zhang, and Y. Wang, Robust axion insulator and Chern insulator phases in a two-dimensional antiferromagnetic topological insulator, Nat. Mater. **19**, 522 (2020).
[27] S. Yang et al., Odd-Even Layer-Number Effect and Layer-Dependent Magnetic Phase Diagrams in MnBi2Te4, Phys. Rev. X **11**, 011003 (2021).
[28] Z. Zhang, N. Wang, N. Cao, A. Wang, X. Zhou, K. Watanabe, T. Taniguchi, B. Yan, and W. Gao, Controlled large non-reciprocal charge transport in an intrinsic magnetic topological insulator MnBi2Te4, Nat Commun **13**, 6191 (2022).
[29] Y. Li et al., Fabrication-induced even-odd discrepancy of magnetotransport in few-layer MnBi2Te4, Nat Commun **15**, 3399 (2024).
[30] N. Liu, S. Schreyeck, K. M. Fijalkowski, M. Kamp, K. Brunner, C. Gould, and L. W. Molenkamp, Antiferromagnetic order in MnBi2Te4 films grown on Si(111) by molecular beam epitaxy, Journal of Crystal Growth **591**, 126677 (2022).
[31] B. Chen et al., Even–Odd Layer-Dependent Exchange Bias Effect in MnBi2Te4 Chern Insulator Devices, Nano Lett. **24**, 8320 (2024).
[32] S.-K. Bac et al., Topological response of the anomalous Hall effect in MnBi2Te4 due to magnetic canting, Npj Quantum Mater. **7**, 46 (2022).
[33] M. M. Otrokov et al., Prediction and observation of an antiferromagnetic topological insulator, Nature **576**, 416 (2019).
[34] S. H. Lee et al., Spin scattering and noncollinear spin structure-induced intrinsic anomalous Hall effect in antiferromagnetic topological insulator MnBi2Te4, Phys. Rev. Research **1**, 012011 (2019).
[35] P. M. Sass, J. Kim, D. Vanderbilt, J. Yan, and W. Wu, Robust A -Type Order and Spin-Flop Transition on the Surface of the Antiferromagnetic Topological Insulator MnBi 2 Te 4, Phys. Rev. Lett. **125**, 037201 (2020).
[36] Z. Lian et al., Antiferromagnetic quantum anomalous Hall effect under spin flips and flops, Nature **641**, 70 (2025).
[37] D. Ovchinnikov et al., Intertwined Topological and Magnetic Orders in Atomically Thin Chern Insulator MnBi2Te4, Nano Lett. **21**, 2544 (2021).
[38] Y.-F. Zhao et al., Even–Odd Layer-Dependent Anomalous Hall Effect in Topological Magnet MnBi2Te4 Thin Films, Nano Lett. **21**, 7691 (2021).
[39] Y. Bai et al., Quantized anomalous Hall resistivity achieved in molecular beam epitaxy-grown MnBi2Te4 thin films, Natl Sci Rev **11**, nwad189 (2024).
[40] L. Tai et al., Distinguishing the Two-Component Anomalous Hall Effect from the Topological Hall Effect, ACS Nano **16**, 17336 (2022).
[41] Y. Shi et al., Correlation between magnetic domain structures and quantum anomalous Hall effect in epitaxial MnBi2Te4 thin films, Phys. Rev. Materials **8**, 124202 (2024).

[42]     F. Hou et al., Te-Vacancy-Induced Surface Collapse and Reconstruction in Antiferromagnetic Topological Insulator MnBi2Te4, ACS Nano **14**, 11262 (2020).
[43]     F. Lüpke, M. Kolmer, J. Yan, H. Chang, P. Vilmercati, H. H. Weitering, W. Ko, and A.-P. Li, Anti-site defect-induced disorder in compensated topological magnet MnBi2-xSbxTe4, Commun Mater **4**, 1 (2023).
[44]     J. (Jens) Als-Nielsen, Des. McMorrow, *Elements of Modern X-Ray Physics*, Wiley, **2011**.
[45]     P. F. Miceli, *Semiconductor Interfaces, Mirostructures and Devices: Properties and Applications*, IOP Publishing, Bristol, **1993**.

**Data Availability**

The data is available from the authors upon reasonable request.

**Supplementary Materials**

See Supplemental Material at [URL will be inserted by publisher] for details on X-ray diffraction simulations and additional transport data which includes addition references [44-45].

**Acknowledgments**

This work was supported by the U. S. Department of Energy (DOE), Office of Science, Basic Energy Sciences (BES), Materials Sciences and Engineering Division (growth, X-ray work and modeling, and transport measurements). The National Quantum Information Science Research Centers, Quantum Science Center (Transport), The NNSA's Laboratory Directed Research and Development Program at Los Alamos National Laboratory (assistance with the X-ray diffraction and modeling). Los Alamos National Laboratory is managed by Triad National Security, LLC for the U.S. Department of Energy's NNSA, under contract 89233218CNA000001. For the electron microscopy work, G.A.V.-L. and D.R.H. acknowledge support through startup funds from the Penn State Eberly College of Science, Department of Chemistry, College of Earth and Mineral Sciences, Department of Materials Science and Engineering, and Materials Research Institute. G.A.V.-L. acknowledges support from the DOE Office of Science, Office of Workforce Development for Teachers and Scientists, Office of Science Graduate Student Research (SCGSR) program. The SCGSR program is administered by the Oak Ridge Institute for Science and Education (ORISE) for the DOE. ORISE is managed by ORAU under contract number DE- SC0014664. Electron microscopy was conducted as part of a user project at the Center for Nanophase Materials Sciences (CNMS), which is a US Department of Energy, Office of Science User Facility at Oak Ridge National Laboratory. L.W. is sponsored by the Army Research Office under Grant Number W911NF-20-2-0166 and W911NF-25-20016.

# Supplemental Material

**Scanning transmission electron microscopy methods**

The scanning transmission electron microscopy (STEM) experiment was performed on a NION UltraSTEM operated at 200kV. The sample was prepared using a focused ion beam system where the sample was milled and then lifted out. Following this, the sample was milled further using Ar at a low energy of 900eV in a Fischione NanoMill.

**X-ray diffraction simulations**

The specular Bragg reflections were simulated using a 1-D kinematic model, which was written in terms of out-of-plane scattering vector $q_z = 4\pi\sin(2\theta/2)/\lambda$ where the wavelength was chosen to be copper $k_{\alpha 1}$ radiation, $\lambda = 1.5406$ Å. For an epitaxial thin film, the specular reflectivity and Bragg diffraction is given by the square modulus of the sum of amplitudes from the individual layers. The intensity $I_{Spec.}$ can be calculated as a function of $q_z$ by

$$I_{Spec.}(q_z) = \frac{S}{q_z^2} \left| \frac{F_{Sub}(q_z)}{1 - e^{-iq_z c_{Sub}}} + \sum_k^n (A_{MBT,1,k}(q_z) + e^{iq_z(t_{MBT,1})}A_{BT,k}(q_z) + e^{iq_z(t_{MBT,1}+t_{BT})}A_{MBT,2,k}(q_z))e^{iq_z(t_k)} \right|^2, \quad (1)$$

$$F_{Sub}(q_z) = \sum_{j=1}^{6} \rho_O e^{iq_z(\frac{1}{12}+\frac{1}{6}j)c_{Sub}} + \rho_{Al} e^{iq_z(0.0188+\frac{1}{6}j)c_{Sub}}, \quad (2)$$

and

$$A_{BT}(q_z) = \sum_{j=0}^{n-1} \rho_{Bi}\left(e^{iq_z(j+0.18)d_{BT}} + e^{iq_z(j+0.56)d_{BT}}\right)e^{-\sigma_{j,Bi}^2 q_z^2} + \rho_{Te}(e^{iq_z(j)d_{BT}} + e^{iq_z(j+0.37)d_{BT}} + e^{iq_z(j+0.74)d_{BT}})e^{-\sigma_{j,Te}^2 q_z^2} \quad (3)$$

$$A_{MBT}(q_z) = \sum_{j=0}^{n-1} \rho_{Bi}\left(e^{iq_z(j+0.13)d_{MBT}} + e^{iq_z(j+0.68)d_{MBT}}\right)e^{-\sigma_{j,Bi}^2 q_z^2} + \rho_{Te}(e^{iq_z(j)d_{MBT}} + e^{iq_z(j+0.29)d_{MBT}} + e^{iq_z(j+0.52)d_{MBT}} + e^{iq_z(j+0.80)d_{MBT}})e^{-\sigma_{j,Te}^2 q_z^2} + \rho_{Mn}\left(e^{iq_z(j+0.40)d_{MBT}}\right)e^{-\sigma_{j,Bi}^2 q_z^2} \quad (4)$$

Here the $F_{Sub}(q_z)$ is the structure factor for the sapphire substrate and $A_{MBT,m}(q_z)$ and $A_{BiTe,m}(q_z)$ for the MnBi$_2$Te$_4$ and Bi$_2$Te$_3$ layers. Here, the summations are over the atoms in each repeat unit along the out-of-plane axis ($z$). The exponential terms of the form $e^{iqz(j+z0)}$ contain the relative position of each atom within the layer, $z_0$, which can be read off from the known structures. $S$ is an overall scale factor, the term in the denominator for the Al$_2$O$_3$ term accounts for the surface truncation, the phase terms of the form $e^{iq_z t}$ models the spatial separation among layers from the substrate interface of the particular layer. In eq. (2) for Al$_2$O$_3$ $c_{Sub} = 12.991$ Å is the lattice parameter along the $c$-axis and the terms in the parenthesis in the exponent index the atomic positions within the unit cell. For eq. (2-4) $\rho_i$ is the atomic weight as an approximation to the atomic form factor. In eq (3) $d_{MBT} = 13.5$ Å and in eq (4) $d_{BT} = 10.0$ Å. Finally, the exponential $e^{-\sigma_k^2 q_z^2}$ is the Debye-Waller term that was included to account for vertical fluctuations in the atomic positions that appears as a fall off of the intensity with increasing $q_z$ and only changes the relative intensities of the reflections. The semiempirical parameters $\sigma_{j,Bi} = 0.001$, $\sigma_{j,Te} = 0.01$ and $\sigma_{j,Mn} = 0.05$ is related to the root-mean-square vertical displacements of atoms[1,2].

The Mn-Bi-Te system is composed of the seven-atom septuple layers (SLs) of $MnBi_2Te_4$ and five-atom quintuple layers (QLs) of $Bi_2Te_3$. Thus, to model this, eq (1) contains a sum over the three layers composed of $MnBi_2Te_4$, $Bi_2Te_3$, and a final layer of $MnBi_2Te_4$, where the thickness of each is then encoded in the summing the individual layers in eq (3) and (4) with the proper phase offset for each layer. This enables simulating several important aspects of the Mn-Bi-Te system. (i) Setting the $Bi_2Te_3$ and the second $MnBi_2Te_4$ layer to zero results in the specular diffraction of pure $MnBi_2Te_4$, which is shown at the bottom of Fig. S1. (ii) Taking the first $MnBi_2Te_4$ to be a single SL, the $Bi_2Te_3$ to be a single QL and second $MnBi_2Te_4$ to be zero results in $MnBi_4Te_7$, shown at the top of Fig. S1. (iii) This model can be used to calculate the structures that are between $MnBi_2Te_4$ and $MnBi_4Te_7$, which can be observed experimentally in Ref. [3]. This is achieved by letting the number of SLs in the first layer be greater than one, a single QL, and the second $MnBi_2Te_4$ layer set to zero, then repeating over this unit. These results are shown in Fig. S1 between the end members. (iv) Important for this study is the case of a single QL intergrowth, which is achieved by setting the $Bi_2Te_3$ layer to be 1 QL and fixing the net thickness of the first and second $MnBi_2Te_4$ layers. For example, as shown in the schematic in Fig. 2(a) of the main text, for a film of 5 SLs, the first layer could be 1 SL and the second layer would then be 4 SLs, giving a stack of SL-QL-SL-SL-SL-SL. The calculations shown in Figure 2 of the main text were obtained by summing the calculations where the QL occupied all possible positions, which were then averaged. For the 5QL example: $|QL\text{-}SL\text{-}SL\text{-}SL\text{-}SL\text{-}SL|^2$ + $|SL\text{-}QL\text{-}SL\text{-}SL\text{-}SL\text{-}SL|^2$ +… $|SL\text{-}SL\text{-}SL\text{-}SL\text{-}SL\text{-}QL|^2$. This then simulates a sample where the formation of a QL occurs randomly across the sample, but with domains larger than the coherence of the X-ray beam, thus not accounting for any coherent effects.

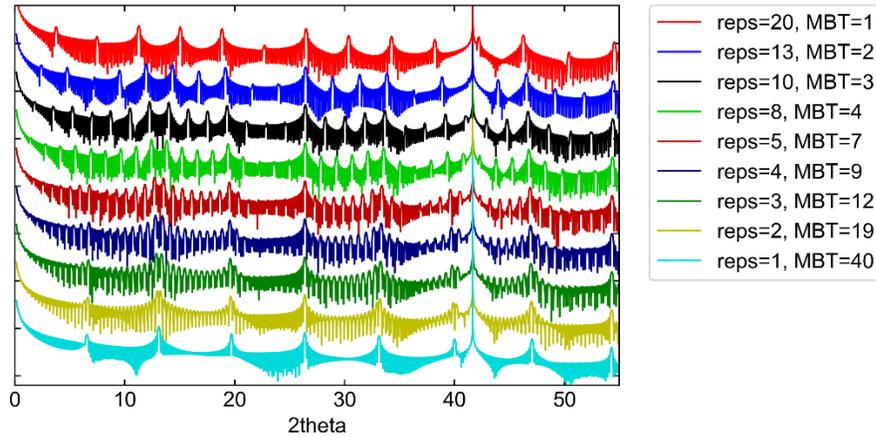

**Fig. S1** Simulated X-ray diffraction curves for samples $MnBi_2Te_4$ (bottom) to $MnBi_4Te_7$ (top), which is achieved by setting the number of $MnBi_2Te_4$ SL in the first layer to the number in the legend, the second $MnBi_2Te_4$ to zero, with a single QL of $Bi_2Te_3$ and the number of repetitions set to the number in the legend. This keeps the nominal thickness fixed at around 50 nm.

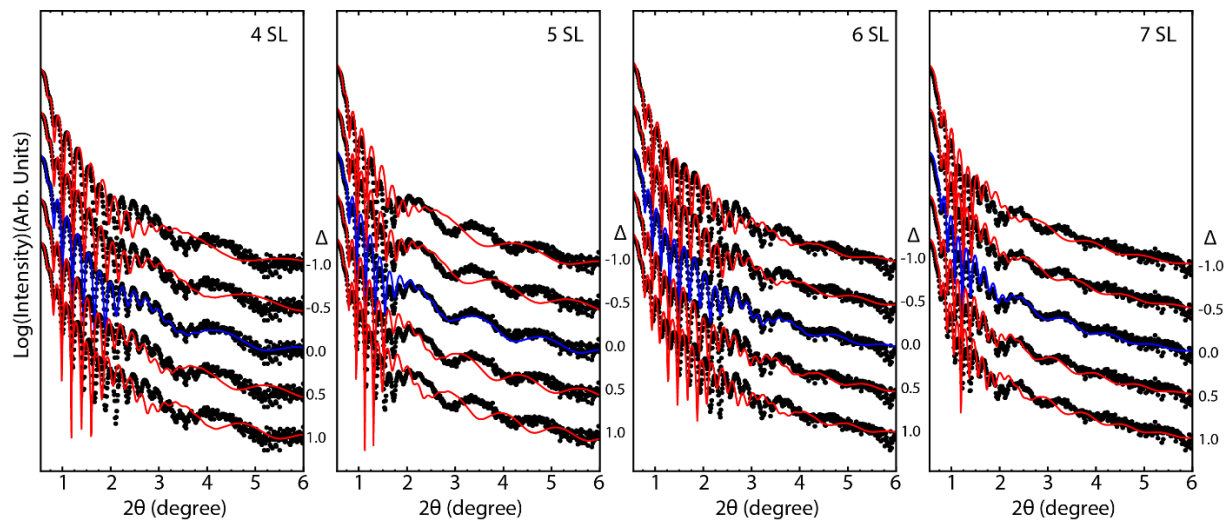

**Fig. S2** X-ray reflectivity data (symbols) and corresponding fits (solid lines) for various thicknesses ($t_0$), labeled at the top. (see Tab. 1 of the main text for fitting parameters). Shown are the best fits (blue curves in the middle), and the additional curves are the reflectivity calculated with different thicknesses, $t_0 + \Delta$, where $\Delta$ = -1.0 SL, -0.5 SL, 0.5 SL, and 1.0 SL, as labeled. Here, the blue curves with $\Delta = 0$ are excellent fits, while the red curves with $\Delta \neq 0$ clearly do not describe the data. This plot shows the near-subangstrom level of sensitivity of X-ray reflectivity to small changes in the thickness, thus demonstrating that X-ray reflectivity is an excellent probe to accurately measure the thickness.

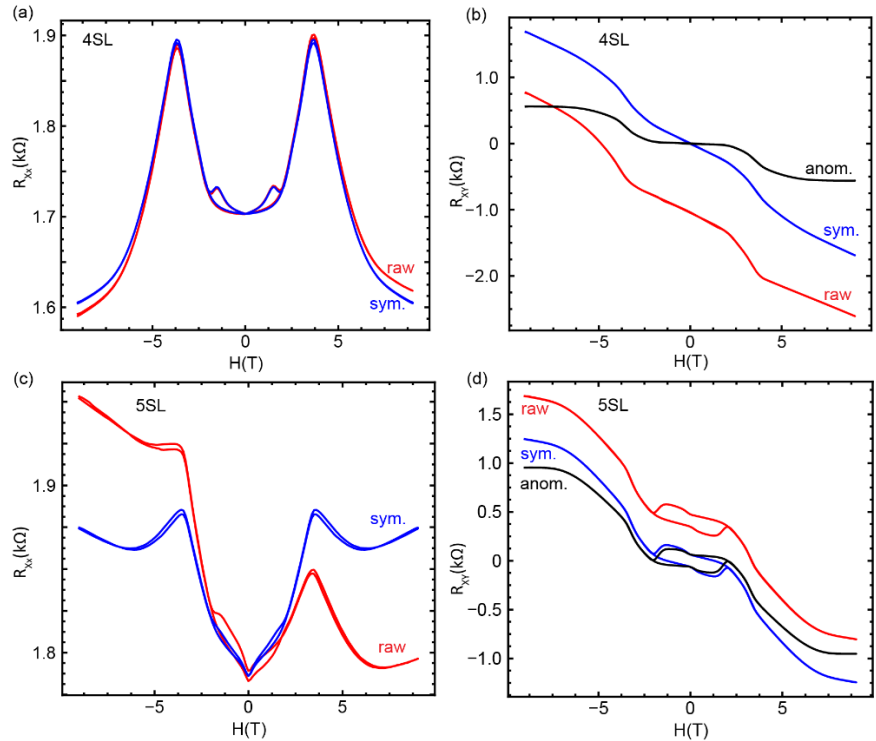

**Fig. S3** Comparison of raw and processed transport data. (a,c) Resistance versus magnetic field for 4 SL (a) and 5 SL (c). Red is the raw data and blue is the symmetrized data, as labeled. (b,d) Hall resistance versus magnetic field for 4 SL (b) and 5 SL (d). Red is the raw data, blue is the symmetrized data and black is the anomalous Hall effect data where a linear background was subtracted from the Hall effect data (slope was fitted from the data in the field range of 9-10T), as labeled.

**References:**

[1] J. (Jens) Als-Nielsen, Des. McMorrow, *Elements of Modern X-Ray Physics*, Wiley, **2011**.
[2] P. F. Miceli, *Semiconductor Interfaces, Mirostructures and Devices: Properties and Applications*, IOP Publishing, Bristol, **1993**.
[3] J. Lapano, L. Nuckols, A. R. Mazza, J. Pai, Yun-Yi, Zhang, B. Lawrie, R. G. Moore, G. Eres, H. N. Lee, M.-H. H. Du, T. Z. Ward, J. S. Lee, W. J. Weber, Y. Zhang, M. Brahlek, *Phys. Rev. Mater.* **2020**, *4*, 111201.